# REMAGNETIZATION PROCESS OF FE-RICH AMORPHOUS WIRE UNDER TIME DEPENDENT TENSILE STRESS

Przemysław Gawroński[1], Alexander Chizhik[3], Juan Mari Blanco[2] and Julian Gonzalez[3]

[1]Faculty of Physics and Applied Computer Science, AGH University of Science and Technology, 30-059, Cracow, Poland
[2]Departamento de Fisica Aplicada I, UPV/EHU, 20080, San Sebastian, Spain
[3]Departamento de Fisica de Materiales, UPV/EHU, 20080, San Sebastian, Spain

An influence of a time dependent external tensile stress on the remagnetization process of a bistable wire is investigated experimentally and numerically. The dependence of the magnetization on the frequency of the tensile stress is analyzed in terms of the magnetic domain structure of Fe-rich wires. The dependence of the switching field on the phase of the tensile stress is presented and reproduced by a straightforward numerical simulation of the remagnetization process in a bistable wire. There, the input is the experimental static dependences of the switching field and of the magnetization of the wire on an external tensile stress. The presented dependences of the switching field and the magnetization on the periodic tensile stress allow to tune the shape of the hysteresis loop, what can be useful in constructing a phase sensor.

**Keywords**: remagnetization process, hysteresis, tensile stress, amorphous wires, bistability

## 1. INTRODUCTION

Recent interest in amorphous magnetic wires is due to their existing and potential application in magnetic sensors[1]. Magnetic properties of amorphous wires strongly depend on their magnetic structure, determined by the stress frozen during the fabrication process. In the magnetic structure of such wires two components can be identified, namely the axially oriented monodomain inner core and the radially oriented polydomain outer shell. The appearance of a single large Barkhausen jump between two stable states of the inner core is a characteristic feature of the remagnetization process of the bistable wires. The remagnetization of the monodomain inner core starts by depinning of the domain wall when the applied magnetic field surpasses the value of the switching field $H_s$, and then follows the propagation of the domain wall along the wire. The switching mechanism is not yet fully understood, especially the exact relation between the stress applied during the preparation and the value of the switching field of the fabricated wire remains unknown. Since we cannot fully predict and control the value of the switching field of the produced wire, we try to tailor the value of $H_s$ a posteriori by applying the external sources as circular magnetic field [2] or tensile stress. The application of the external tensile stress during the remagnetization process causes the monotonic growth of the remanent magnetization and non monotonic modification of the switching field. This variation of the switching field is due to the competition between the growth of the volume of the inner core and a modification of the closure domains at the wire ends as the external tensile stress grows [3,4].

In this paper we study the remagnetization process of the Fe-rich amorphous wires in the presence of a time-dependent external tensile stress, being the positive part of the sinusoidal function. The paper is organized as follows. In two subsequent sections we present the results of our experimental studies and the results of the simulation based on a straightforward theoretical model. Final conclusions close the text.

## 2. MEASUREMENT

The axial hysteresis loops of amorphous wire of nominal composition $Fe_{77.5}B_{15}Si_{7.5}$ were measured in the presence of the time-variable external tensile stress. The wire was placed inside the sine-like magnetic field produced by Helmholtz coils, the frequency of the applied field was equal to $v=20Hz$ and the amplitude to $H_m=12\ A/m$. The length of the wire was *8cm* and the diameter *125μm*. In order to apply the time-variable tensile stress one end of the wire was fixed to the sample holder, while the other one was attached to the mechanical vibrator. The amplitude and the frequency of the tensile stress were controlled through the power supply connected to the mechanical vibrator.

In Figs. 1,2 we present the time dependence of the magnetization and the remanence of a single wire obtained for the applied tensile stress of the amplitude equal to $S_{ts}=400MPa$ and the frequency $f = 0.5Hz$. The presented time dependence of the magnetization can be divided into two regimes. In the first regime, marked in the figure as *A*, the tensile stress is absent. The hysteresis loop for this part is presented in the inset *a*.

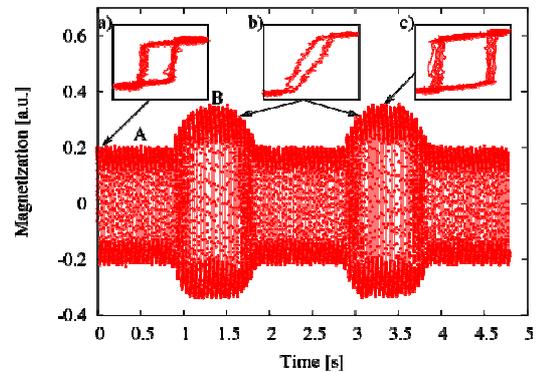

**Fig. 1.** The effect of time dependent tensile stress ($S_{ts}=400MPa$, $f=0.5Hz$) on the magnetization of a single wire. The insets show the modification of hysteresis loops under the time dependent stress.

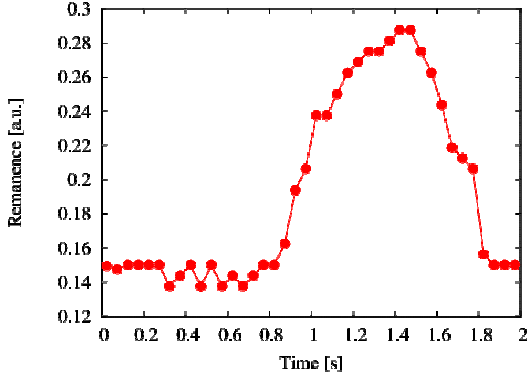

**Fig. 2**. The remanence as a function of time due to changes in applied time dependent stress ($S_{ts}=400MPa, f=0.5Hz$).

The hysteresis loop is squared, and in all the subsequent cycles in this regime, the magnetization switches between two constant values. In the regime marked in the figure as *B*, the changes of the magnetization in the subsequent cycles follows the increase and the decrease of the tensile stress. As the value of the tensile stress grows in the subsequents cycles of the applied magnetic field the wire switches between increasing values of the magnetization. The growing external stress causes the inner core to extend what manifests itself as the growth of the magnetization. On the other hand rather drastic fall of $H_s$ form *5A/m* to almost *0A/m* observed in insets *a* and *b* in Figure 1 is probably due to the growth of the closure domains as the tensile stress increases [3,4]. It is worth to notify that the squareness of the hysteresis loop is also lost in the inset *b*, as well as the bistability. The inset *c* presents the hysteresis loop obtained for the maximal value of the applied stress. There, the magnetization is maximal, as well as the switching field [3,4]. The changes of $H_s$ are supposed to be the result of the competition between the shrinking of the closure domains and an expansion of the inner core. For the hysteresis loop presented in the inset *c* the latter process dominates.

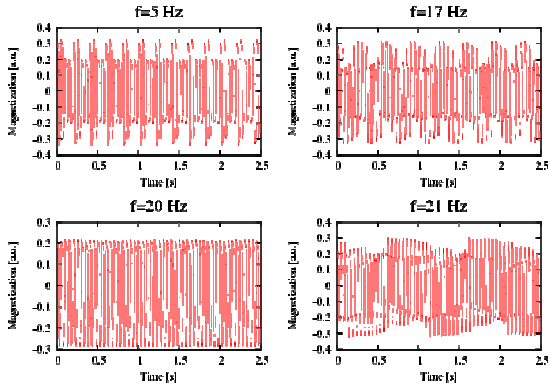

**Fig. 3.** The influence of the frequency of the time dependent tensile stress on the magnetization of a single wire.

Figure 3 shows the time dependence of the magnetization for the amplitude of the tensile stress equal to *400MPa* and for various frequencies *f* of the tensile stress. The presented time profiles of the magnetization suggest that the frequency of the applied tensile stress could be used as a main driving source of the time changes of the magnetization of the sample. If the two frequencies *v* and *f* are equal or *f/v* ratio is a natural number, then the remanent magnetization in subsequent cycles of the applied field is constant, e.g. *f=20Hz*, in Figure 3. The magnetic response of the wire depends on the two periodic signals, the applied magnetic field and the applied tensile stress. In order to study the influence of the relation between these two signals on the remagnetization process we used the power supply with two outputs. One output was connected to the Helmholtz coils generating the applied magnetic field and the second output was connected to the mechanical vibrator producing the external tensile stress. Such a setup allows us to investigate the influence of the phase $\varphi_{ts}$ of the tensile stress on the remagnetization process. The amplitude of the applied magnetic field was set to *12A/m*, the amplitude of the tensile stress was $S_{ts}=80MPa$. The frequency of the applied field was *v=10Hz*. The frequency of the tensile stress was *f=k\*v*, where k=*1,2,3,4,5,6*.

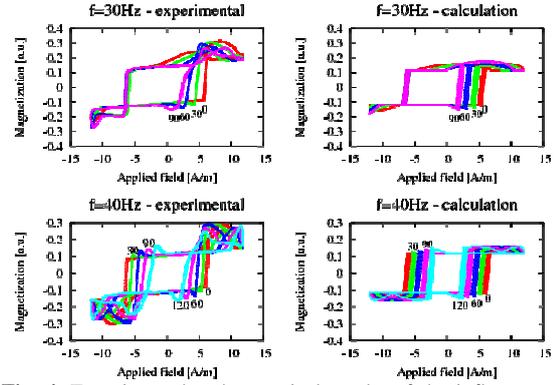

**Fig. 4.** Experimental and numerical results of the influence of the phase of the time dependent tensile stress ($S_{ts}=80MPa$) on the shape of the hysteresis loops of a single wire.

In Figure 4 we present the sample hysteresis loops for two frequencies of the tensile stress, namely *f=30Hz* and *f=40Hz* and various phases $\varphi_{ts}$ of the tensile stress. In the hysteresis loop for a single wire two switching field values are to be distinguished; the positive switching field $H_{s+}$ is the value of the applied field when the wire magnetized down flips up, and negative $H_{s-}$ when the wire flips down. Let us consider the experimental hysteresis loops obtained for the *f=30Hz*, as an example of an odd ratio of *f/v*, presented in top left in Figure 4. The increase of the value of the phase $\varphi_{ts}$ of the tensile stress from zero to *90deg* results in the decrease of $H_{s+}$ from *6A/m* to *2A/m*. $H_{s-}$ is almost not affected by the change of the phase $\varphi_{ts}$. The experimental hysteresis loops for case of even ratio of *f/v (f=40Hz)* are presented in the bottom left of Figure 4. In this case the absolute values of the positive and the negative switching field decrease from *6A/m* to *1A/m* while the phase $\varphi_{ts}$ increases from zero to *120deg*. The variation of $H_s$ with the phase of the tensile stress $\varphi_{ts}$ for two frequencies of the tensile stress, namely *f=30Hz* and *f=40Hz* are presented in Figure 5. For the case of an odd frequencies *f/v* ratio e.g. *f=30Hz* (Figure 5 top)*,* an increase of the phase of the stress from zero to *150deg* leads $H_{s+}$ to decrease, leaving $H_{s-}$ almost constant. On the contrary, when the phase decreases from *-20* to *-180deg* the negative switching field $H_{s-}$ decreases, leaving $H_{s+}$ almost constant. Figure 5 bottom represents the behavior of $H_s$ in the case when the ratio of *f/v* is even, e.g. *f=40Hz*. This time both $H_{s+}$ and $H_{s-}$ decrease as the phase of the stress increases from zero to *150deg*. The phases of the stress from zero to about *-150deg* nearly do not affect any of $H_s$, and then for *-150deg* an abrupt change of $H_s$ occurs.

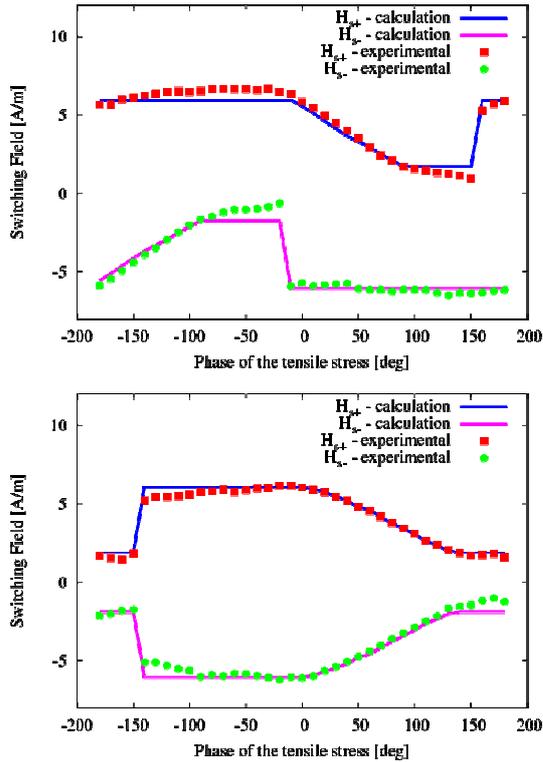

**Fig. 5.** The experimental and numerical dependences of the switching field on the phase of the tensile stress of the amplitude of *80MPa* and frequency *f=30Hz* (top) and *f=40Hz* (bottom).

## 3. NUMERICAL CALCULATION

In the case of the absence of the external tensile stress the simulation scheme of the hysteresis loop of a bistable wire is straightforward. If we neglect the finite velocity of the domain wall, the simulated bistable wire flips immediately its magnetization, as soon as the applied magnetic field reaches the value of the switching field $H_s$. Since we know from the experimental works [3] the static dependence of $H_s$ and the magnetization on the tensile stress, we can incorporate them in the simulation procedure in the following way.

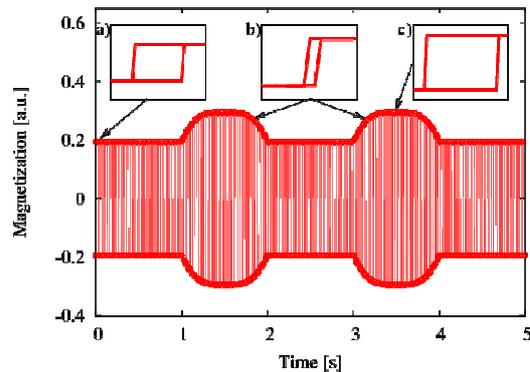

**Fig. 6**. The numerical calculation of the of time dependent tensile stress ($S_{ts}=400MPa$, $f=0.5Hz$) on the magnetization on a single wire. The insets show the modification of hysteresis loops under the time dependent stress.

The time dependence of the tensile stress is a positive part of the sinusoidal function $S(t)=S_{ts} sin(2\pi ft+ \varphi_{ts})$. We iterated the applied field $H_{app}(t)=H_m sin(2\pi vt)$ and the tensile stress at the same time with the step e. g. $dt=10^{-3} s$, and for each value of stress we calculated the actual values of $H_s(S(t))$ and $M(S(t))$ from the fitted static experimental dependences. The switching of the magnetization took place when $H_{app}(t) \geq H_s(S(t))$. In Figure 6 we present the calculated time dependence of the magnetization of a single wire taking into account the time dependent stress of the amplitude equal to $S_{ts}=400MPa$ and the frequency *f=0.5Hz*. The hysteresis loops presented in the insets *a, b, c* in Figure 6 are the numerical reproductions of the experimental hysteresis loops from insets of Figure 1. The numerical simulation was also preformed in order to study the dependence of $H_s$ on the phase of the tensile stress $\varphi_{ts}$. The numerically calculated hysteresis loops for the same parameters as in the case of the measurement are presented in the right side of Figure 4. The simulated hysteresis loops reproduce the behavior of the measured ones. For the case of *f=30Hz*, the calculated positive switching field decreases with the increase of the phase of the tensile stress. For the case of *f=40Hz* the positive and the negative switching fields decrease with the increase of $\varphi_{ts}$. Solid lines in Figure 5 represents the simulated dependence of $H_s$ on the phase $\varphi_{ts}$ of the tensile stress. There is a good agreement between the simulated and experimental values.

## 4. CONCLUSIONS

The most important goal of this paper is that the static experimental dependences of switching field and magnetization on the tensile stress are sufficient inputs to reproduce numerically the experimental dynamic dependences of the magnetization and the switching field on the external tensile stress. This conclusion can be drawn from the comparison of Figures 1 and 6 and of the plots at Figures 4 and 5. The application of a time-dependent stress gives us a new possibility to smoothly and reversibly adjust the magnetic properties of Fe-rich bistable wire, such as the switching field and the remanent magnetization. The rate of changes of the remagnetization process can be controlled via the frequency and the amplitude of the applied stress. Samples with time dependent magnetic properties can be of potentially use in sensor applications. The shape of the hysteresis loop is also sensitive to the phase of the stress, if the ratio of *f/v* is a natural number. Since the phase of the stress controls the value of $H_s$, this relationship can be useful in constructing a phase sensor.

**Acknowledgements**
This work was supported by the „Krakow Interdisciplinary Ph.D.-Project in Nanoscience and Advances Nanostructures" operated within Foundation for Polish Science MPD Programme cofinanced by European Regional Development Fund.